%% file: main.tex
\documentclass[a4paper]{article}

\usepackage{INTERSPEECH2021}
\usepackage{microtype}
\usepackage{graphicx}

\usepackage{subcaption}
\usepackage{caption}
\usepackage{amssymb}
\usepackage{amsmath}
\usepackage{booktabs} 
\usepackage{xcolor} 
\usepackage{pgfplots,pgfplotstable}

\usepackage{hyperref}

\usepackage{siunitx}

\usepackage{enumitem}

\usepackage{cite}

\setlist{nosep} 

\usepackage[colorinlistoftodos]{todonotes}

\newcommand\slimeq{\mkern1.5mu{=}\mkern1.5mu}

\definecolor{forestgreen}{rgb}{0.13, 0.55, 0.13}
\definecolor{fulvous}{rgb}{0.86, 0.52, 0.0}
\definecolor{glaucous}{rgb}{0.38, 0.51, 0.71}
\definecolor{lava}{rgb}{0.81, 0.06, 0.13}
\definecolor{buff}{rgb}{0.94, 0.86, 0.51}
\definecolor{chromeyellow}{rgb}{1.0, 0.65, 0.0}
\definecolor{brightube}{rgb}{0.82, 0.62, 0.91}

\pgfplotsset{scaled y ticks=false, scaled x ticks=false}

\pgfplotstableread[col sep=comma]{data/mcd/flowtron_base_mcd.csv.csv}\ftabasemcd
\pgfplotstableread[col sep=comma]{data/mcd/flowtron_baseline_mcd.csv.csv}\ftabaselinemcd
\pgfplotstableread[col sep=comma]{data/mcd/flowtron_ctc_mcd.csv.csv}\ftactcmcd

\pgfplotstableread[col sep=comma]{data/mcd/tacotron_baseline_mcd.csv.csv}\tacotronbaselinemcd
\pgfplotstableread[col sep=comma]{data/mcd/tacotron_ctc_mcd.csv.csv}\tacotronctcmcd
\pgfplotstableread[col sep=comma]{data/mcd/tacotron_without_prior_mcd.csv.csv}\tacotronbasewopriormcd

\pgfplotstableread[col sep=comma]{data/mcd/fastpitch_grapheme_baseline_mcd.csv}\fastpitchbaselinemcd
\pgfplotstableread[col sep=comma]{data/mcd/fastpitch_grapheme_mcd.csv}\fastpitchconvmcd

\pgfplotstableread[col sep=comma]{data/mcd/fastspeech2_mfa_mcd.csv}\fastspeechbaseline
\pgfplotstableread[col sep=comma]{data/mcd/fastspeech2_convattn_mcd.csv}\fastspeechconvattn

\pgfplotstableread[col sep=comma]{data/mcd/collated_with_betab_mcd_dtw.csv}\ftpbetabinomialmcddtw
\pgfplotstablesort[sort key=iteration, sort cmp=int >]{\ftpbetabinomialmcddtwsorted}{\ftpbetabinomialmcddtw}
\pgfplotstableread[col sep=comma]{data/mcd/collated_with_mfa_mcd_dtw.csv}\ftpmfamcddtw
\pgfplotstablesort[sort key=iteration, sort cmp=int >]{\ftpmfadtwsorted}{\ftpmfamcddtw}
\pgfplotstableread[col sep=comma]{data/mcd/collated_without_prior_mcd_dtw.csv}\ftpnopriormcddtw
\pgfplotstablesort[sort key=iteration, sort cmp=int >]{\ftpnopriormcddtwsorted}{\ftpnopriormcddtw}

\pgfplotstableread[col sep=comma]{data/duration/distance_flowtron_baseline.csv}\ftabasedurationdistance
\pgfplotstableread[col sep=comma]{data/duration/distance_flowtron_CTC_guided.csv}\ftactcdurationdistance

\pgfplotstableread[col sep=comma]{data/cer/memfix_fta_base.csv}\ftacerbase
\pgfplotstableread[col sep=comma]{data/cer/memfix_fta_ctc.csv}\ftacerctc
\pgfplotstableread[col sep=comma]{data/cer/taco2_base_smoothed-v2.csv}\tacocerbase
\pgfplotstableread[col sep=comma]{data/cer/taco2_ctc_smoothed-v2.csv}\tacocerctc
\pgfplotstableread[col sep=comma]{data/cer/ftp_dp_smoothed-v2.csv}\ftpcer

\pgfplotstableread[col sep=comma]{data/duration/distance_tacotron_base.csv}\tacotronbasedurationdistance
\pgfplotstableread[col sep=comma]{data/duration/distance_tacotron_ctc.csv}\tacotronctcdurationdistance

\newcommand{\attn}[1]{\mathcal{A}_{\mathit{#1}}}

\title{One TTS Alignment To Rule Them All}
\name{Rohan Badlani, Adrian Łańcucki, Kevin J. Shih, Rafael Valle, Wei Ping, Bryan Catanzaro}

\address{
  NVIDIA 
}

\email{\{rbadlani,alancucki,kshih,rafaelvalle,wping,bcatanzaro\}@nvidia.com}

\begin{document}

\maketitle

\begin{abstract}
    Speech-to-text alignment is a critical component of neural text-to-speech (TTS) models. Autoregressive TTS models typically use an attention mechanism to learn these alignments on-line. However, these alignments tend to be brittle and often fail to generalize to long utterances and out-of-domain text, leading to missing or repeating words. Most non-autoregressive end-to-end TTS models rely on durations extracted from external sources. In this paper we leverage the alignment mechanism proposed in RAD-TTS as a generic alignment learning framework, easily applicable to a variety of neural TTS models. The framework combines forward-sum algorithm, the Viterbi algorithm, and a simple and efficient static prior.
    In our experiments, the alignment learning framework improves all tested TTS architectures, both autoregressive (Flowtron, Tacotron 2) and non-autoregressive (FastPitch, FastSpeech 2, RAD-TTS). Specifically, it improves alignment convergence speed of existing attention-based mechanisms, simplifies the training pipeline, and makes the models more robust to errors on long utterances.
    Most importantly, the framework improves the perceived speech synthesis quality, as judged by human evaluators.
\end{abstract}


\input{intro.tex}
\input{method.tex}
\input{experiments.tex}
\input{conclusion.tex}

\pagebreak
\bibliographystyle{IEEEtran}
\bibliography{main}

\end{document}

%% file: intro.tex
\section{Introduction}

Neural text-to-speech (TTS) models, especially autoregressive TTS models, produce naturally sounding speech for in-domain text~\cite{Tacotron, Tacotron2, Flowtron}.
However, these models can suffer from pronunciation issues such as missing and repeated words for out-of-domain text, especially in long utterances. A typical neural TTS model consists of an encoder that maps text inputs to hidden states, a decoder that generates mel-spectograms or waveforms from the hidden states, and an alignment mechanism or a duration source that maps the encoder states to decoder inputs~\cite{Tacotron, Tacotron2, Flowtron, fastspeech2, yiFastSpeech, kim2020glowtts, lancucki2020fastpitch}. Autoregressive TTS models rely on the attention mechanism~\cite{Graves13, bahdanau2014neural} to align text and speech, typically using content based attention mechanism~\cite{Tacotron, Flowtron}. Although recent works have improved alignments by using both content and location sensitive attention~\cite{Tacotron2}, such models still suffer from alignment problems on long utterances~\cite{kim2020glowtts}. 

In contrast, parallel (non-autoregressive) TTS models factor out durations from the decoding process, thereby requiring durations as input for each token. These models generally rely on external aligners~\cite{fastspeech2}  like the Montreal Forced Aligner (MFA)~\cite{MFA}, or on durations extracted from a pre-trained autoregressive model (or forced aligner)~\cite{lancucki2020fastpitch, peng2020non, yiFastSpeech} like Tacotron 2~\cite{Tacotron2}. In addition to the dependency on external alignments, these models 
can suffer from poor training efficiency, 
require carefully engineered training schedules to prevent unstable learning, and may be difficult to extend to languages either because pre-existing aligners are either unavailable or their output does not exactly fit the desired format. Ideally, we would like the alignment to be trained end-to-end as part of the TTS model to significantly simplify the training pipeline. We would also like the alignments to converge and stabilize rapidly as the rest of the TTS pipeline is dependent on it. Most importantly the output quality should be no worse (and hopefully better) than if we were to train on alignments provided by external sources.


 This work leverages the alignment framework proposed in~\cite{rad_tts} to simplify alignment learning in several TTS models. We demonstrate its ability to convert all TTS models to a simpler end-to-end pipeline with better convergence rates and improved robustness to long utterances. We improve prior work on alignments in autoregressive TTS systems~\cite{Tacotron, Tacotron2, Flowtron} by adding a constraint that directly maximizes the likelihood of text given speech mel-spectrograms. We demonstrate that this approach can also be used to learn alignments online in parallel TTS models~\cite{lancucki2020fastpitch, fastspeech2, rad_tts}, again eliminating the need for external aligners or alignments obtained from a pre-trained TTS models. In addition, we further examine the effect of a simple, static alignment prior for guiding alignment attention learning \cite{tachibana2018efficiently, rad_tts}.
 We demonstrate in our experiments that our framework can improve \emph{both} autoregressive and parallel models with respect to convergence rate of speech text alignments, closeness to hand-annotated durations, and speech quality. In summary, our results\footnote{Samples available at \hyperlink{https://nv-adlr.github.io/one-tts-alignment}{https://nv-adlr.github.io/one-tts-alignment}} show that TTS models trained with our alignment learning framework have fewer repeated and missing words during inference, improved stability on long sequence synthesis, and improved overall speech quality based on human evaluation.

%% file: method.tex
\section{Alignment Learning Framework}
\begin{figure*}[!t]
    \centering
    \includegraphics[width=0.9\linewidth]{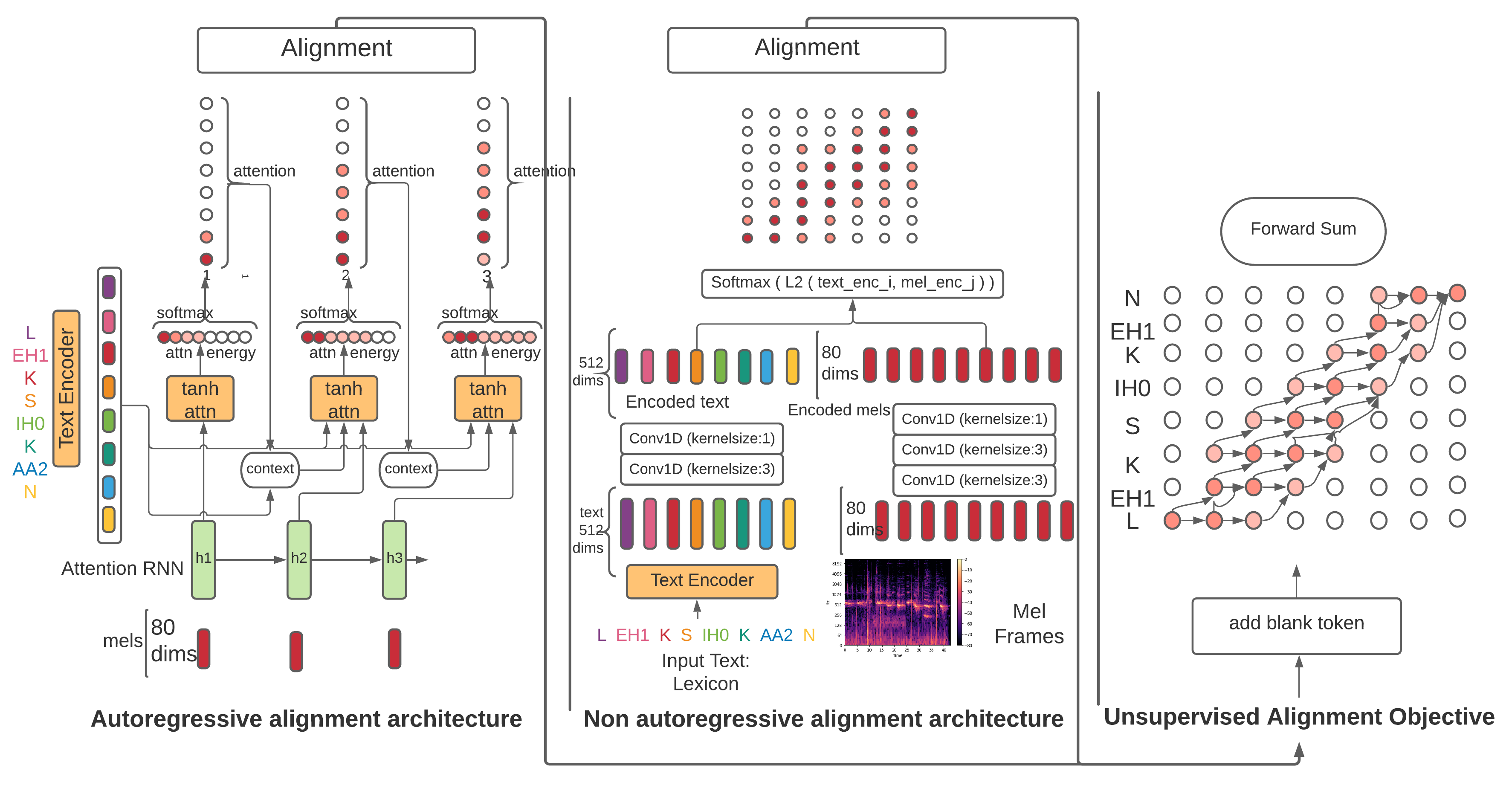}
    \vspace{-1em}
    \caption{Overview of our Alignment Learning Framework: autoregressive models use a sequential attention mechanism to generate alignments between text and mels. Non-autoregressive models encode text and mels using simple 1D convolutions and use pairwise $L_2$ distance to compute the alignments. The alignments represent the distribution $P(s_t | x_t)$ and the alignment objective (Equation \ref{eq:likelihood}).}
    \label{fig:alignment_arch}
    \vspace*{-0.5\baselineskip}
\end{figure*}

We extend the alignment learning approach proposed in RAD-TTS~\cite{rad_tts} to be more broadly applicable to various text to speech models especially autoregressive models. Our alignment framework is presented in Figure~\ref{fig:alignment_arch}. It takes the encoded text input $\Phi \in \mathbb{R}^{C_{\mathit{txt}} \times N}$ and aligns it to mel-spectrograms  $X \in \mathbb{R}^{C_{\mathit{mel}} \times T}$ where $T$ is number of mel frames and $N$ is the text length. In this section, we introduce the alignment learning objective and its application to autoregressive and parallel models.

\subsection{Unsupervised alignment learning objective}
\label{sec:ctc_objective}
To learn the alignment between mel-spectrograms $X$ and text $\Phi$, we use the alignment learning objective proposed in RAD-TTS~\cite{rad_tts}. This objective maximizes the likelihood of text given mel-spectrograms using the forward-sum algorithm used in Hidden Markov
Models (HMMs)~\cite{Rabiner89-ATO}. In our formulation, we constrain the alignment between text and speech to be monotonic, in order to avoid missing or repeating tokens. The following equation 
summarizes the likelihood of text given mels~\cite{rad_tts}:
%
%
\begin{align}\label{eq:likelihood}
 P \left(S(\Phi) \mid X;\theta\right) = \sum_{\mathbf{s} \in S(\Phi)} {\prod_{t=1}^{T}{P(s_t \mid x_t;\theta)} }
\end{align}
where $s$ is a specific alignment between mel-spectrograms and text (eg: $s1 = \phi_1, s2 = \phi_1, s3 = \phi_2, \ldots, sT = \phi_N$), $S(\Phi)$ is the set of all possible valid monotonic alignments, $ P(s_t | x_t)$ is the likelihood of a specific text token $s_t = \phi_i$ aligned for mel frame $x_t$ at timestep $t$. It is important to note that the above formulation of the alignment learning objective does not depend on how the likelihood  $P(s_t = \phi_i \mid x_t)$ is obtained. Hence, it can be applied to both autoregressive and parallel models. We define the forward sum objective that maximizes (\ref{eq:likelihood}) as $\mathcal{L}_{\mathit{ForwardSum}}$. 
Following RAD-TTS, we use an efficient, off-shelf CTC~\cite{ctc_icml2006} implementation to compute this objective (details in appendix of RAD-TTS~\cite{rad_tts}). 

\subsection{Autoregressive TTS Models}
\label{sec:autoregressive_alignment_framework}
Autoregressive TTS models typically use a sequential formulation of attention to learn online alignments. TTS models such as Tacotron~\cite{Tacotron} and Flowtron~\cite{Flowtron} use a content based attention mechanism that relies only on decoder inputs and the current attention hidden state to compute an attention map between encoder and decoder steps. Other autoregressive models use a location relative attention mechanism~\cite{locationattn} to promote forward movement of alignments~\cite{Tacotron2}. Although alignment learning in these autoregressive models is tightly coupled with the decoder and can be learned with the mel-spectrogram reconstruction objective, it has been observed that the likelihood of a misstep in the alignment increases with the length of the utterance. This results in catastrophic failure on long sequences and out-of-domain text~\cite{Battenberg}. The application of the unsupervised objective described in Sec \ref{sec:ctc_objective} improves both convergence speed during training and robustness during inference. 

Our autoregressive setup uses the standard stateful content based attention mechanism for Flowtron~\cite{Flowtron} and a hybrid attention mechanism that uses both content and location based features for Tacotron2~\cite{Tacotron2}. 
The location sensitive term (Eq. \ref{eq:location_term}) uses features computed from attention weights at previous decoder timesteps. 
We use a Tacotron2 encoder to obtain the sequence of encoded text representations $(\phi_i^{enc})_{i=1}^{N}$ and an attention RNN to produce a sequence of states $h_t$. A simple architecture is used to compute the alignment energies $e_{t, i}$ for text token $s_i$ at timestep $t$ for mel $x_t$ using the tanh attention~\cite{bahdanau2014neural}. The attention weights are computed with softmax over the text domain using the alignment energies. The following equations summarize the attention mechanism:
\begin{equation}
  \begin{aligned}
     (h_{{t}})_{t=1}^{T} = \text{RNN}(h_{t-1}, x_{t-1}, c_{t-1})
   \end{aligned}
\end{equation}
\begin{equation}
  \begin{aligned}
     c_{t} = \sum \alpha_{t, i}  \phi^{enc}_i 
   \end{aligned}
\end{equation}
\begin{equation}\label{eq:location_term}
  \begin{aligned}
     f_t = F (\alpha_{t-1})
   \end{aligned}
\end{equation}
\begin{equation}
  \begin{aligned}
     e_{t, i} = -v^{T} \tanh ( Wh_{t} + V\phi_i^{enc} + Uf_{t, i})
   \end{aligned}
\end{equation}
\begin{equation}
  \begin{aligned}
     P(s_t = \phi_i | x_t)= \alpha_{t, i} = Softmax(- e_t)_i,
   \end{aligned}
\end{equation}
where $f_t$ is the location relative term for location sensitive attention $F$ (cumulative attention from~\cite{Tacotron2} using a concatenation of the attention weights from the previous timestep and the cumulative attention weights). 
The attention weights model the distribution $P(s_t = \phi_i | x_t)$, which is exactly the right-most term in Equation (\ref{eq:likelihood}), and we incorporate it as the alignment loss:
\begin{equation}
  \begin{aligned}
     \mathcal{L}_{\mathit{align}} = \mathcal{L}_{\mathit{ForwardSum}}.
   \end{aligned}
\end{equation}


\subsection{Parallel TTS Models}
\label{sec:non_autoregressive_alignment_framework}
As parallel TTS models have durations factored out from the decoder, the alignment learning module can be decoupled from the mel decoder as a standalone aligner. This provides a lot of flexibility in choosing the architecture to formulate the distribution $ P(s_t | x_t)$, where $s_t$ is a random variable for a text token aligned at timestep $t$ for mel frame $x_t$. Similar to GlowTTS~\cite{kim2020glowtts} and RAD-TTS~\cite{rad_tts}, we compute the soft alignment distribution based on the learned pairwise affinity between all text tokens and mel frames, which is normalized with softmax across the text domain

\begin{align}
    &D_{i,j} = dist_{L2}(\phi_i^{enc}, x_j^{enc}),\\
    &\attn{soft} = \texttt{softmax}(-D, \texttt{dim}=0).
\end{align}

We use two simple convolutional encoders from RAD-TTS~\cite{rad_tts} for encoding text $\Phi$ as $\Phi^{enc}$ and mel-spectograms $X$ as $X^{enc}$ with 2 and 3 1D convolution layers respectively. In Section \ref{sec:experiments}, we demonstrate that the same architecture works well with different parallel TTS models such as FastPitch and FastSpeech 2. Parallel models require alignments to be specified beforehand, typically in the form of the number of output samples for every input phoneme, equivalent to a binary alignment map. However, attention models produce soft alignment maps, constituting a train-test domain gap. Following~\cite{rad_tts,kim2020glowtts}, we use the Viterbi algorithm to find the most likely monotonic path through the soft alignment map in order to convert soft alignments ($\attn{soft}$) to hard alignments ($\attn{hard}$). We further close the gap between soft and hard alignments by forcing $\attn{soft}$ to match $\attn{hard}$ as much as possible by minimizing their KL-divergence. This is used in both Glow-TTS and RAD-TTS, formulated as ${L}_{bin}$:
\begin{equation}
    \mathcal{L}_{bin} = \attn{hard} \odot \log\attn{soft},
\end{equation}
\begin{equation}
  \begin{aligned}
     \mathcal{L}_{\mathit{align}} = \mathcal{L}_{\mathit{ForwardSum}} + \mathcal{L}_{bin}.
   \end{aligned}
\end{equation}
where $\odot$ is Hadamard product, ${L}_{\mathit{align}}$ is final alignment loss.

 
\subsection{Alignment Acceleration}
\label{sec:accelerating_alignments}
Faster convergence of alignments means faster training for the full TTS model, as the decoder needs a stable alignment representation to build upon. During training, since the length of mel-spectrograms is known, we use a static 2D prior~\cite{rad_tts}, that is wider near the center and narrower near the corners to accelerate the alignment. This idea has been previously explored by Tachibana et al~\cite{guidedloss} where they introduce a new loss promoting near-diagonal alignments. Although our formulation with the 2D static prior is slightly different than Tachibana et al~\cite{guidedloss}, but we believe both should yield similar results. The 2D prior substantially accelerates the alignment learning by making far-off-diagonal elements less probable, although other priors can also be used for this goal. We apply this prior $f_{B}$ over the alignment $P(s \mid X \slimeq x_t)$ to obtain the following posterior:
 \begin{multline}
     f_{B}(k, \alpha, \beta) = {N \choose k} \frac{B(k + \alpha)B(N-k+\beta)}{B(\alpha, \beta)} 
 \end{multline}
 \begin{multline}
 P_{posterior}(\Phi \slimeq \phi_{k} \mid X\slimeq x_t)=\\
     P(\Phi \slimeq \phi_{k} \mid X\slimeq x_t) \odot f_{B}(k, \omega  t,  \omega  (T-t+1))
\end{multline}
 for $k=\{0,\ldots,N\}$, where $\alpha$, $\beta$ are hyperparameters of beta function $B(\cdot, \cdot)$, $N$ is number of tokens and $\omega$ is scaling factor controlling width of prior: lower the $\omega$, wider the width. 

%% file: experiments.tex
\section{Experiments}
\label{sec:experiments}
We evaluate the effectiveness of the alignment learning framework by comparing its performance in terms of convergence speed, distance from human annotated ground truth durations, and speech quality. For autoregressive models like Flowtron and Tacotron 2, we compare with the baseline alignment methods therein. For FastPitch, we compare with an alignment method that relies on an external TTS model (Tacotron2) to obtain token durations. For the parallel models: FastSpeech 2 and RAD-TTS, we compare against an alignment method that obtains durations from the MFA aligner. We use the LJ Speech dataset (LJ)~\cite{LJS} for all our experiments. 

\subsection{Convergence Rate}
\begin{figure}[h]
    \begin{subfigure}{\linewidth}
    \label{fig:taco2_mcd}
    \hfill\input{tacotron_mcd_plot.tex}\hspace{1.0cm}
 \end{subfigure}
 \begin{subfigure}[t]{\linewidth}
     \label{fig:fta_mcd}
     \hfill\input{fta_mcd_plot.tex}\hspace{1.0cm}
 \end{subfigure}
 \begin{subfigure}[t]{\linewidth}
    \label{fig:rad_tts_mcd}
    \hfill\input{ftp_mcd_plot.tex}\hspace{1.0cm}
 \end{subfigure}
  \begin{subfigure}[t]{\linewidth}
    \label{fig:fastpitch_mcd}
     \hfill\input{fastpitch_mcd_plot.tex}\hspace{1.0cm}
 \end{subfigure}
 \begin{subfigure}[t]{\linewidth}
    \label{fig:fastspeech2_mcd}
    \hfill\input{fastspeech2_mcd_plot.tex}\hspace{1.0cm}
 \end{subfigure}
 \caption{Convergence rate improvements in TTS models with the alignment learning framework}
 \label{fig:mcd}
 \vspace*{-1\baselineskip}
\end{figure}
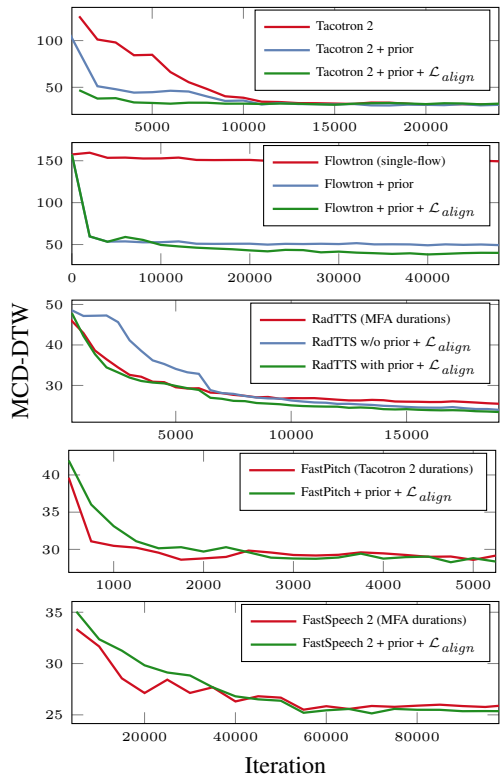

In order to compare the convergence rate of different alignment methods, we use the mean mel-cepstral distance (MCD)~\cite{mcd, Battenberg}. MCD compares the distance between synthesized and ground truth mel-spectrograms aligned temporarily with dynamic time warping (DTW). We observe in Figure~\ref{fig:mcd} that using the static prior described in Section~\ref{sec:accelerating_alignments} significantly improves the convergence rate of Tacotron2. Parallel models such as RAD-TTS, FastPitch, and FastSpeech2 with the alignment framework (no dependency on external aligners) converge at the same rate as their baseline models using a forced aligner. The model that benefits the most from using the alignment framework is Flowtron. It has two autoregressive flows running in opposing directions, each with their own learned alignment. Notably, the second autoregressive flow is performed on top of the autoregressive outputs of the previous flow. This means that if the alignment in the first flow fails, so will the second. Training is very slow as the second flow can only be added after the first has converged. Prior attempts to train both flows simultaneously have resulted in poor minima where neither flow has learned to align. By using just the attention prior, we are now able to train at least two flows simultaneously, with further improvements with adding the unsupervised alignment learning $\mathcal{L}_{\mathit{align}}$ objective described in Section~\ref{sec:ctc_objective}. This significantly reduces training time and improves convergence of Flowtron. 


\begin{figure*}
 \begin{subfigure}{0.25\textwidth}
    \centering
    \includegraphics[width=\linewidth]{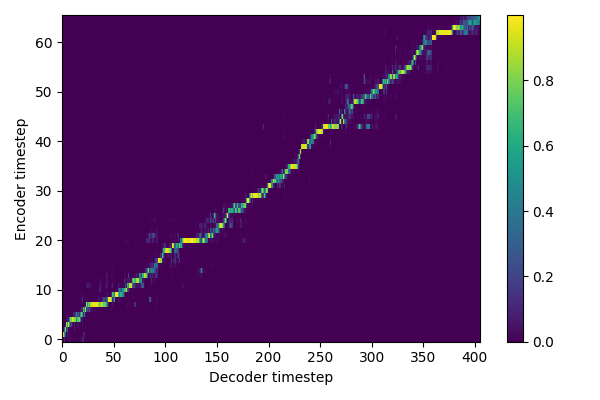}
    \caption{Flowtron}
    \label{fig:fta_base_align}
\end{subfigure}
\begin{subfigure}{0.25\textwidth}
\centering
  \includegraphics[width=\linewidth]{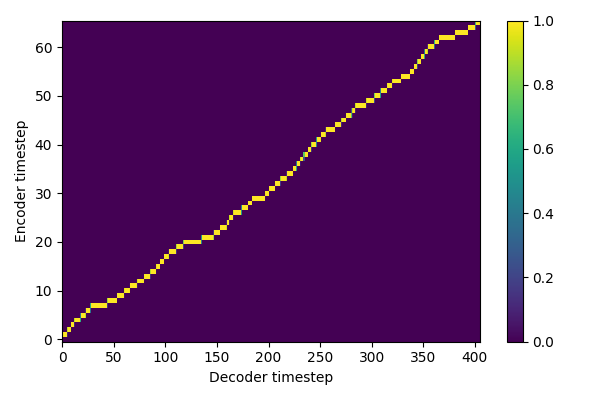}
  \caption{Flowtron with $\mathcal{L}_{\mathit{align}}$}
  \label{fig:fta_ctc_align}
\end{subfigure}
 \begin{subfigure}{0.25\textwidth}
 \centering
  \includegraphics[width=\linewidth]{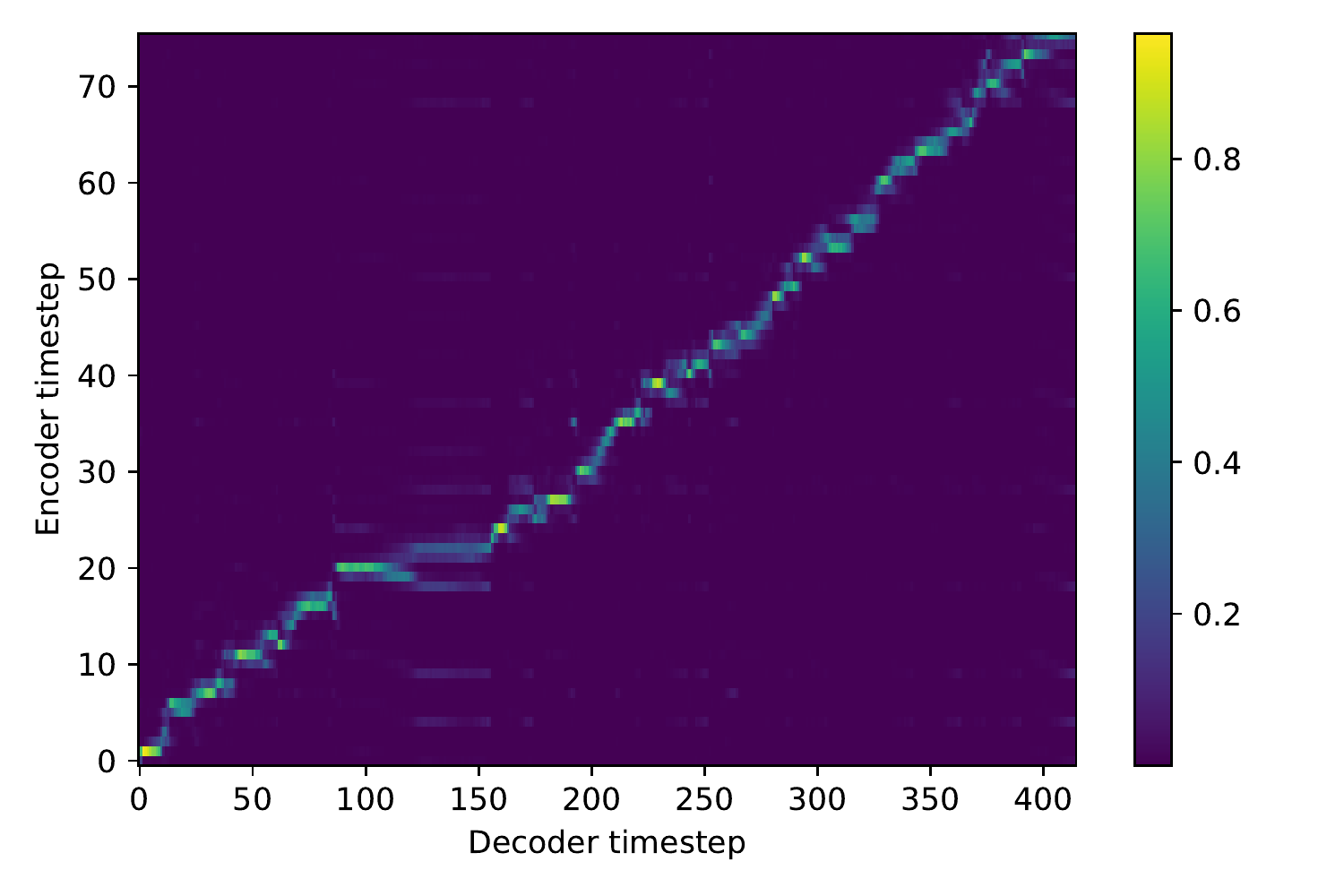}
  \caption{Tacotron 2}
  \label{fig:taco2_base_align}
\end{subfigure}%
\begin{subfigure}{0.25\textwidth}
\centering
  \includegraphics[width=\linewidth]{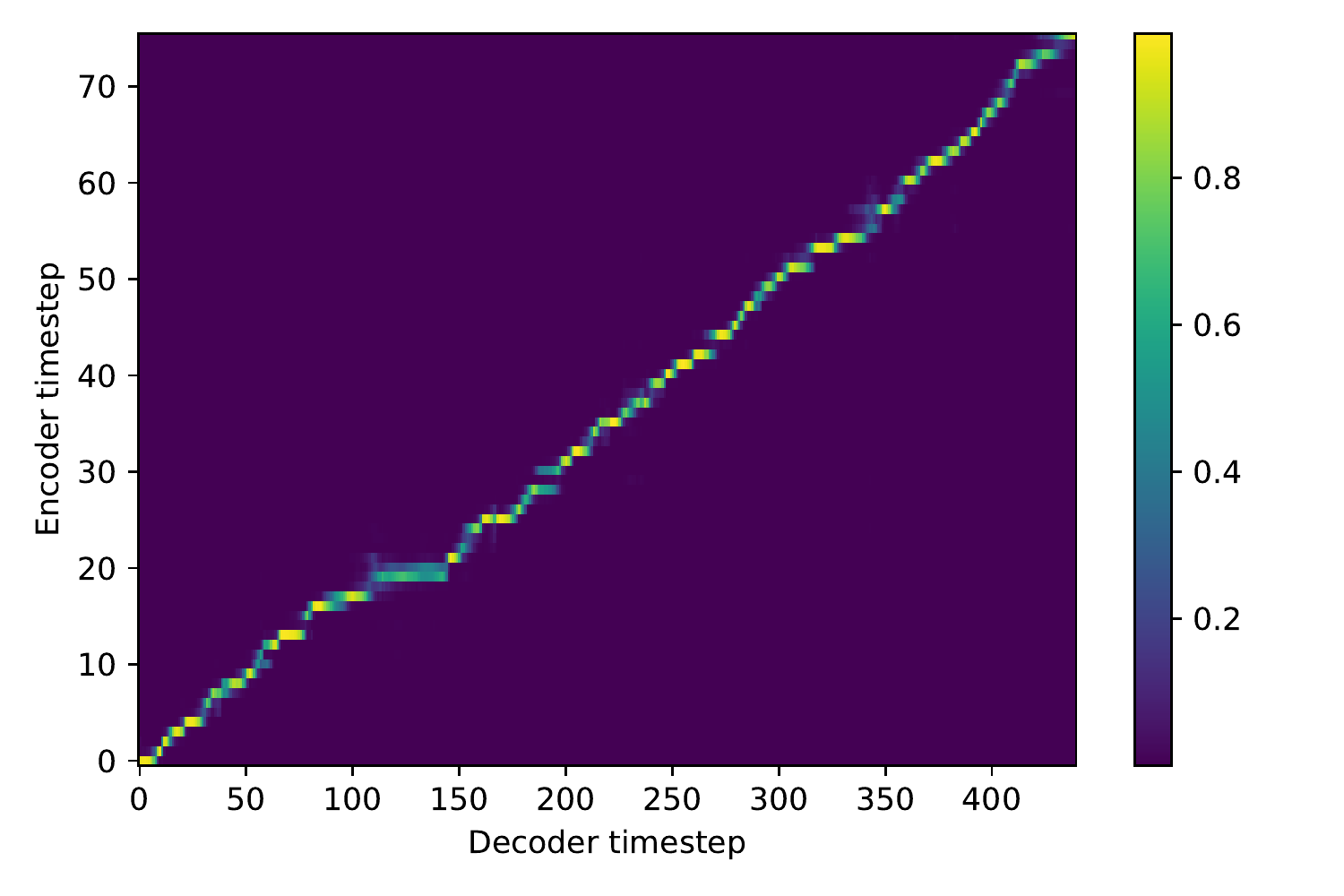}
  \caption{Tacotron 2 with $\mathcal{L}_{\mathit{align}}$}
  \label{fig:taco2_ctc_align}
\end{subfigure}
\caption{Converged soft alignments for Flowtron, Tacotron2. Alignment framework provides sharper and more connected alignments.}
\label{fig:align_convergence}
\vspace*{-1.5\baselineskip}
\end{figure*}

\subsection{Alignment Sharpness}
We visually inspect alignment matrices for a specific validation sample in Figure~\ref{fig:align_convergence}. The alignment objective consistently makes the attention distribution sharper with more connected alignment paths. This suggests that models with $\mathcal{L}_{\mathit{align}}$ produce more confident and continuous alignments, and by extension, continuous speech without repeating or missing words. 

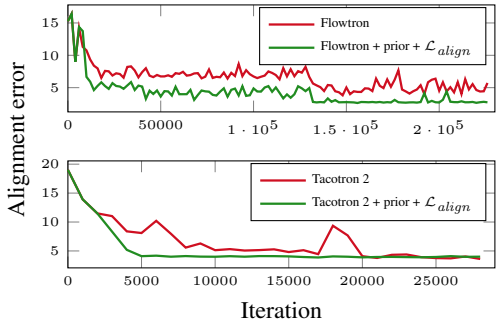
\begin{figure}[h]
\begin{subfigure}{\linewidth}
    \label{fig:flowtron_duration_diff}
    \hfill\input{fta_duration_distance_plot.tex}\hspace{1.0cm}
 \end{subfigure}
 \begin{subfigure}{\linewidth}
    \label{fig:tacotron_duration_diff}
    \vspace{-1.2cm}  
    \hfill\input{tacotron_duration_distance_plot.tex}\hspace{1.0cm}
 \end{subfigure}
 \vspace{-1\baselineskip}
 \caption{$L_1$ distance between ground truth alignments and those extracted during training for Flowtron and Tacotron 2. Both use different batch sizes and are thus plotted separately. 
 }
 \label{fig:duration_diff}
 \vspace*{-1.5\baselineskip}
\end{figure}

\subsection{Duration Analysis}
In order to observe the influence of the unsupervised alignment loss on the quality of alignments, we compare phoneme durations extracted from model alignments to manually annotated phoneme durations from 10 samples of the LJ test set. For autoregressive models, we extract the binarized alignments from soft alignments using a monotonic argmax, iterating through phonemes and identifying the phoneme with maximum attention weights among the current and next phonemes. We use this binarized alignment to extract durations for each phoneme. Figure~\ref{fig:duration_diff} shows the average $L_1$ distance between durations extracted from the models with respect to ground truth annotated durations. By using our alignment framework we obtain a faster convergence rate than the baseline and alignments closer to the ground truth.


\subsection{Pairwise Opinion Scores}
We crowd-sourced pairwise preference scores to subjectively compare models trained with our alignment learning framework against baseline. Listeners were pre-screened with a hearing test based on sinusoid counting. During the pairwise ranking, raters were repeatedly given two synthesized utterances of the same text, picked at random from 100 LJ test samples. Both were synthesized with the same architecture: one being the baseline, and other using our alignment framework. The listeners were shown the text and asked to select samples with the best overall quality, defined by accuracy of text, its pleasantness and naturalness. Approximately 200 scores per model were collected. Table~\ref{tab:pairwise_results} shows pairwise preference scores of models trained with alignment framework over baseline. It shows that the alignment framework consistently improves over all baselines.

\begin{table}[!ht]
    \caption{Pairwise preference scores judged by human raters, shown with $95\%$ confidence intervals. Scores above 0.5 indicate models trained with $\mathcal{L}_{\mathit{align}}$ were preferred by majority of raters.}
    \centering
    \begin{tabular}{lc}
        \toprule
        \textbf{Model} & \textbf{ Alignment Framework vs Baseline}\\
        \midrule
        Tacotron 2  & $0.556 \pm 0.068$ \\
        Flowtron $(\sigma = .5)$ & $0.635 \pm 0.065$ \\
        RAD-TTS $(\sigma = .5)$ & $0.639 \pm 0.066$ \\
        FastPitch & $0.565 \pm 0.068$ \\
        FastSpeech2 & $0.521 \pm 0.067$ \\
        \bottomrule
        \end{tabular}
    \vspace{-1.5\baselineskip}
    \label{tab:pairwise_results}
    
\end{table}
\subsection{Robustness to Errors on Long Utterances}
\begin{figure}[h]
    
 \begin{subfigure}{\linewidth}
    \input{taco2_long_utterance.tex}
    \vspace{-1.0\baselineskip}
 \end{subfigure}
 \caption{Character error rate of different models at different text lengths. Models that use the alignment framework make fewer mistakes with increased utterance length.}
 \label{fig:cer}
 \vspace*{-1.5\baselineskip}
\end{figure}
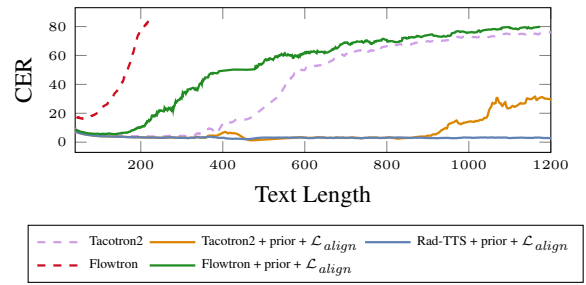

We measure character error rate (CER) between synthesized and input texts using an external speech recognition model to evaluate the robustness of the alignments on long utterances. We use $14,045$ full sentences from the LibriTTS dataset~\cite{libritts}. We synthesize speech with models trained on LJ Speech, and recognize it with Jasper~\cite{li2019jasper}. Figure~\ref{fig:cer} shows that autoregressive models with $\mathcal{L}_{\mathit{align}}$ have a lower CER, providing evidence that the alignment objective results in more robust speech for long utterances. Parallel models such as RAD-TTS use a duration predictor and do not suffer from alignment issues, and hence have a much lower CER than autoregressive models. 


%% file: tacotron_mcd_plot.tex
\begin{tikzpicture}[]
  \begin{axis}[
    height=3.0cm,
    width=0.9\linewidth,
      legend cell align={left},
      ylabel near ticks,
      xlabel near ticks,
      xlabel style= {font=\small},
      ylabel style={font=\small},
      legend style={font=\tiny},
      yticklabel style = {font=\tiny},
      xticklabel style = {font=\tiny},
      xmin=600,
      xmax=24000,
      x tick label style={/pgf/number format/1000 sep=},  
      legend columns=1,
      legend entries={Tacotron 2, Tacotron 2 + prior, Tacotron 2 + prior + $\mathcal{L}_{\mathit{align}}$}
    ]
    \addplot [line width=0.3mm, color=lava] table [x=model, y=mel_cepstrum_mcd, y error=oursstd,col sep=comma] {\tacotronbasewopriormcd};
    \addplot [line width=0.3mm, color=glaucous] table [x=model, y=mel_cepstrum_mcd, y error=oursstd,col sep=comma] {\tacotronbaselinemcd};
    \addplot [line width=0.3mm, color=forestgreen] table [x=model, y=mel_cepstrum_mcd, y error=oursstd,col sep=comma] {\tacotronctcmcd};
  \end{axis}
\end{tikzpicture}

%% file: fta_mcd_plot.tex
\begin{tikzpicture}[]
  \begin{axis}[
    height=3.2cm,
    width=0.9\linewidth,
      legend cell align={left},
      ylabel near ticks,
      xlabel style= {font=\small},
      ylabel style={font=\small},
      legend style={font=\tiny},
      yticklabel style = {font=\tiny},
      xticklabel style = {font=\tiny},
      x tick label style={
		/pgf/number format/1000 sep={}},
     xmin=0,
     xmax=48000,
      legend columns=1,
      legend entries={Flowtron (single-flow), Flowtron + prior, Flowtron + prior + $\mathcal{L}_{\mathit{align}}$ }
    ]
    \addplot [line width=0.3mm, color=lava] table [x=model, y=mel_cepstrum_mcd, y error=oursstd,col sep=comma] {\ftabasemcd};
    \addplot [line width=0.3mm, color=glaucous] table [x=model, y=mel_cepstrum_mcd, y error=oursstd,col sep=comma] {\ftabaselinemcd};
    \addplot [line width=0.3mm, color=forestgreen] table [x=model, y=mel_cepstrum_mcd, y error=oursstd,col sep=comma] {\ftactcmcd};
  \end{axis}
\end{tikzpicture}

%% file: ftp_mcd_plot.tex
\begin{tikzpicture}[]
  \begin{axis}[
    height=3.2cm,
    width=0.9\linewidth,
      legend cell align={left},
      ylabel= {MCD-DTW},
      ylabel near ticks,
      xlabel near ticks,
      xlabel style= {font=\small},
      ylabel style={font=\small},
      legend style={font=\tiny},
      yticklabel style = {font=\tiny},
      xticklabel style = {font=\tiny},
      xmin=500,
      xmax=19000,
      x tick label style={/pgf/number format/1000 sep=},  
      legend columns=1,
      legend entries={RadTTS (MFA durations), RadTTS w/o prior + $\mathcal{L}_{\mathit{align}}$, RadTTS with prior + $\mathcal{L}_{\mathit{align}}$}
    ]
    \addplot [line width=0.3mm, color=lava] table [x=iteration, y=mel_cepstrum_mcd, y error=oursstd,col sep=comma] {\ftpmfadtwsorted};

    \addplot [line width=0.3mm, color=glaucous] table [x=iteration, y=mel_cepstrum_mcd, y error=oursstd,col sep=comma] {\ftpnopriormcddtwsorted};
    
    \addplot [line width=0.3mm, color=forestgreen] table [x=iteration, y=mel_cepstrum_mcd, y error=oursstd,col sep=comma] {\ftpbetabinomialmcddtwsorted};
    
  \end{axis}
\end{tikzpicture}

%% file: fastpitch_mcd_plot.tex
\begin{tikzpicture}[]
  \begin{axis}[
    height=3.2cm,
    width=0.9\linewidth,
      ylabel= { },
      legend cell align={left},
      ylabel near ticks,
      xlabel near ticks,
      xlabel style= {font=\tiny},
      ylabel style={font=\tiny},
      legend style={font=\tiny},
      yticklabel style = {font=\tiny},
     xticklabel style = {font=\tiny},
      xmin=500,
      xmax=5250,
      x tick label style={/pgf/number format/1000 sep=},  
      legend columns=1,
      legend entries={FastPitch (Tacotron 2 durations), FastPitch + prior + $\mathcal{L}_{\mathit{align}}$}
    ]
    \addplot [line width=0.3mm, color=lava] table [x=model, y=mel_cepstrum_mcd, y error=oursstd,col sep=comma] {\fastpitchbaselinemcd};
    \addplot [line width=0.3mm, color=forestgreen] table [x=model, y=mel_cepstrum_mcd, y error=oursstd,col sep=comma] {\fastpitchconvmcd};
  \end{axis}
\end{tikzpicture}

%% file: fastspeech2_mcd_plot.tex
\begin{tikzpicture}[]
  \begin{axis}[
    height=3.2cm,
    width=0.9\linewidth,
      xlabel= Iteration,
      legend cell align={left},
      ylabel near ticks,
      xlabel near ticks,
      xlabel style= {font=\small},
      ylabel style={font=\tiny},
      legend style={font=\tiny},
      yticklabel style = {font=\tiny},
      xticklabel style = {font=\tiny},
      xmin=4000,
      xmax=98000,
      x tick label style={/pgf/number format/1000 sep=},  
      legend columns=1,
      legend entries={FastSpeech 2 (MFA durations), FastSpeech 2 + prior + $\mathcal{L}_{\mathit{align}}$}
    ]
    \addplot [line width=0.3mm, color=lava] table [x=model, y=mel_cepstrum_mcd, y error=oursstd,col sep=comma] {\fastspeechbaseline};
    \addplot [line width=0.3mm, color=forestgreen] table [x=model, y=mel_cepstrum_mcd, y error=oursstd,col sep=comma] {\fastspeechconvattn};
  \end{axis}
\end{tikzpicture}

%% file: fta_duration_distance_plot.tex
\begin{tikzpicture}[scale=1.0]
  \begin{axis}[
  height=3.0cm,
  width=0.9\linewidth,
      legend cell align={left},
      ylabel near ticks,
      xlabel near ticks,
      xlabel style={font=\small},
      ylabel style={font=\small},
      yticklabel style = {font=\tiny},
      xticklabel style = {font=\tiny},
      x tick label style={/pgf/number format/1000 sep=},  
      xmin=0,
      xmax=230000,
      legend style={font=\tiny},
      legend columns=1,
      legend entries={Flowtron, Flowtron + prior + $\mathcal{L}_{\mathit{align}}$}
    ]
    \addplot [line width=0.3mm, color=lava] table [x=iteration, y=phoneme_l1_distance, y error=oursstd,col sep=comma] {\ftabasedurationdistance};
    \addplot [line width=0.3mm, color=forestgreen] table [x=iteration, y=phoneme_l1_distance, y error=oursstd,col sep=comma] {\ftactcdurationdistance};
    
  \end{axis}
\end{tikzpicture}

%% file: tacotron_duration_distance_plot.tex
\begin{tikzpicture}[scale=1.0]
  \begin{axis}[
  height=3.0cm,
  width=0.9\linewidth,
      legend cell align={left},
      xlabel= Iteration,
      ylabel= {Alignment error},
      ylabel near ticks,
      y label style={at={(-0.07,1.2)}},
      xlabel near ticks,
      xlabel style={font=\small},
      ylabel style={font=\small},
      yticklabel style = {font=\tiny},
      xticklabel style = {font=\tiny},
      x tick label style={/pgf/number format/1000 sep=},  
      xmin=0,
      xmax=29000,
      legend style={font=\tiny},
      legend columns=1,
      legend entries={Tacotron 2, Tacotron 2 + prior + $\mathcal{L}_{\mathit{align}}$}
    ]
    \addplot [line width=0.3mm, color=lava] table [x=iteration, y=phoneme_l1_distance, y error=oursstd,col sep=comma] {\tacotronbasedurationdistance};
    \addplot [line width=0.3mm, color=forestgreen] table [x=iteration, y=phoneme_l1_distance, y error=oursstd,col sep=comma] {\tacotronctcdurationdistance};
    
  \end{axis}
\end{tikzpicture}

%% file: taco2_long_utterance.tex
\begin{center}
\begin{tikzpicture}[scale=1.0]
  \begin{axis}[
  height=3.5cm,
  width=0.98\linewidth,
      legend cell align={left},
      xlabel= {Text Length},
      ylabel= {CER},
      ylabel near ticks,
      xlabel near ticks,
      xlabel style={font=\small},
      ylabel style={font=\small},
      yticklabel style = {font=\tiny},
      xticklabel style = {font=\tiny},
      xmin=40,
      xmax=1200,
      x tick label style={/pgf/number format/1000 sep=},  
      legend style={font=\tiny},
      legend style={at={(-0.1,-0.5)},anchor=north west},
      legend columns=3,
      legend entries={Tacotron2, Tacotron2 + prior + $\mathcal{L}_{\mathit{align}}$, Rad-TTS + prior + $\mathcal{L}_{\mathit{align}}$, Flowtron, Flowtron + prior + $\mathcal{L}_{\mathit{align}}$}
    ]
    \addplot [dashed, line width=0.3mm, color=brightube] table [x=text_len, y=mean_cer, y error=oursstd,col sep=comma] {\tacocerbase};
    \addplot [line width=0.3mm, color=fulvous] table [x=text_len, y=mean_cer, y error=oursstd,col sep=comma] {\tacocerctc};
    \addplot [line width=0.3mm, color=glaucous] table [x=text_len, y=mean_cer, y error=oursstd,col sep=comma] {\ftpcer};
    \addplot [dashed, line width=0.3mm, color=lava] table [x=text_len, y=mean_cer, y error=oursstd,col sep=comma] {\ftacerbase};
    \addplot [line width=0.3mm, color=forestgreen] table [x=text_len, y=mean_cer, y error=oursstd,col sep=comma] {\ftacerctc};
    
  \end{axis}
\end{tikzpicture}
\end{center}

%% file: conclusion.tex
\section{Conclusion}
We present an alignment framework that is broadly applicable to various TTS architectures, both autoregressive and parallel. 
By combining proper guidance in the form of forward-sum, Viterbi and diagonal priors, attention-based online alignment learning can be made stable and fast-converging. The alignment learning framework eliminates the need for forced aligners which are expensive to use and often not readily available for certain languages. Our experiments demonstrate improvements in overall speech quality based on human pairwise comparisons, reduced alignment failures, faster convergence, as well as robustness to errors in synthesis of long text sequences.


%% file: main.bbl
\begin{thebibliography}{10}
\providecommand{\url}[1]{#1}
\csname url@samestyle\endcsname
\providecommand{\newblock}{\relax}
\providecommand{\bibinfo}[2]{#2}
\providecommand{\BIBentrySTDinterwordspacing}{\spaceskip=0pt\relax}
\providecommand{\BIBentryALTinterwordstretchfactor}{4}
\providecommand{\BIBentryALTinterwordspacing}{\spaceskip=\fontdimen2\font plus
\BIBentryALTinterwordstretchfactor\fontdimen3\font minus
  \fontdimen4\font\relax}
\providecommand{\BIBforeignlanguage}[2]{{%
\expandafter\ifx\csname l@#1\endcsname\relax
\typeout{** WARNING: IEEEtran.bst: No hyphenation pattern has been}%
\typeout{** loaded for the language `#1'. Using the pattern for}%
\typeout{** the default language instead.}%
\else
\language=\csname l@#1\endcsname
\fi
#2}}
\providecommand{\BIBdecl}{\relax}
\BIBdecl

\bibitem{Tacotron}
\BIBentryALTinterwordspacing
Y.~Wang, R.~J. Skerry{-}Ryan, D.~Stanton, Y.~Wu, R.~J. Weiss, N.~Jaitly,
  Z.~Yang, Y.~Xiao, Z.~Chen, S.~Bengio, Q.~V. Le, Y.~Agiomyrgiannakis,
  R.~Clark, and R.~A. Saurous, ``Tacotron: {A} fully end-to-end text-to-speech
  synthesis model,'' \emph{CoRR}, vol. abs/1703.10135, 2017. [Online].
  Available: \url{http://arxiv.org/abs/1703.10135}
\BIBentrySTDinterwordspacing

\bibitem{Tacotron2}
\BIBentryALTinterwordspacing
J.~Shen, R.~Pang, R.~J. Weiss, M.~Schuster, N.~Jaitly, Z.~Yang, Z.~Chen,
  Y.~Zhang, Y.~Wang, R.~J. Skerry{-}Ryan, R.~A. Saurous, Y.~Agiomyrgiannakis,
  and Y.~Wu, ``Natural {TTS} synthesis by conditioning wavenet on mel
  spectrogram predictions,'' \emph{CoRR}, vol. abs/1712.05884, 2017. [Online].
  Available: \url{http://arxiv.org/abs/1712.05884}
\BIBentrySTDinterwordspacing

\bibitem{Flowtron}
R.~Valle, K.~Shih, R.~Prenger, and B.~Catanzaro, ``Flowtron: an autoregressive
  flow-based generative network for text-to-speech synthesis,'' 2020.

\bibitem{fastspeech2}
Y.~Ren, C.~Hu, T.~Qin, S.~Zhao, Z.~Zhao, and T.-Y. Liu, ``Fastspeech 2: Fast
  and high-quality end-to-end text-to-speech,'' \emph{arXiv preprint
  arXiv:2006.04558}, 2020.

\bibitem{yiFastSpeech}
Y.~Ren, Y.~Ruan, X.~Tan, T.~Qin, S.~Zhao, Z.~Zhao, and T.-Y. Liu, ``Fastspeech:
  Fast, robust and controllable text to speech,'' in \emph{Advances in Neural
  Information Processing Systems}, H.~Wallach, H.~Larochelle, A.~Beygelzimer,
  F.~d\textquotesingle Alch\'{e}-Buc, E.~Fox, and R.~Garnett, Eds.,
  vol.~32.\hskip 1em plus 0.5em minus 0.4em\relax Curran Associates, Inc.,
  2019, pp. 3171--3180.

\bibitem{kim2020glowtts}
J.~Kim, S.~Kim, J.~Kong, and S.~Yoon, ``Glow-tts: A generative flow for
  text-to-speech via monotonic alignment search,'' 2020.

\bibitem{lancucki2020fastpitch}
A.~Łańcucki, ``Fastpitch: Parallel text-to-speech with pitch prediction,''
  2020.

\bibitem{Graves13}
\BIBentryALTinterwordspacing
A.~Graves, ``Generating sequences with recurrent neural networks,''
  \emph{CoRR}, vol. abs/1308.0850, 2013. [Online]. Available:
  \url{http://arxiv.org/abs/1308.0850}
\BIBentrySTDinterwordspacing

\bibitem{bahdanau2014neural}
\BIBentryALTinterwordspacing
D.~Bahdanau, K.~Cho, and Y.~Bengio, ``Neural machine translation by jointly
  learning to align and translate,'' 2014, cite arxiv:1409.0473Comment:
  Accepted at ICLR 2015 as oral presentation. [Online]. Available:
  \url{http://arxiv.org/abs/1409.0473}
\BIBentrySTDinterwordspacing

\bibitem{MFA}
M.~McAuliffe, M.~Socolof, S.~Mihuc, M.~Wagner, and M.~Sonderegger, ``Montreal
  forced aligner: Trainable text-speech alignment using kaldi,'' in
  \emph{INTERSPEECH}, 2017.

\bibitem{peng2020non}
K.~Peng, W.~Ping, Z.~Song, and K.~Zhao, ``Non-autoregressive neural
  text-to-speech,'' in \emph{International Conference on Machine
  Learning}.\hskip 1em plus 0.5em minus 0.4em\relax PMLR, 2020, pp. 7586--7598.

\bibitem{rad_tts}
\BIBentryALTinterwordspacing
K.~J. Shih, R.~Valle, R.~Badlani, A.~Lancucki, W.~Ping, and B.~Catanzaro,
  ``{RAD}-{TTS}: Parallel flow-based {TTS} with robust alignment learning and
  diverse synthesis,'' in \emph{ICML Workshop on Invertible Neural Networks,
  Normalizing Flows, and Explicit Likelihood Models}, 2021. [Online].
  Available: \url{https://openreview.net/forum?id=0NQwnnwAORi}
\BIBentrySTDinterwordspacing

\bibitem{tachibana2018efficiently}
H.~Tachibana, K.~Uenoyama, and S.~Aihara, ``Efficiently trainable
  text-to-speech system based on deep convolutional networks with guided
  attention,'' in \emph{2018 IEEE International Conference on Acoustics, Speech
  and Signal Processing (ICASSP)}.\hskip 1em plus 0.5em minus 0.4em\relax IEEE,
  2018, pp. 4784--4788.

\bibitem{Rabiner89-ATO}
L.~R. Rabiner, ``A tutorial on hidden {Markov} models and selected applications
  in speech recognition,'' \emph{Proceedings of the IEEE}, vol.~77, no.~2, pp.
  257--286, Feb. 1989.

\bibitem{ctc_icml2006}
A.~Graves, S.~Fern\'{a}ndez, F.~Gomez, and J.~Schmidhuber, ``Connectionist
  temporal classification: Labelling unsegmented sequence data with recurrent
  neural networks,'' ser. ICML '06.\hskip 1em plus 0.5em minus 0.4em\relax New
  York, NY, USA: Association for Computing Machinery, 2006, p. 369–376.

\bibitem{locationattn}
\BIBentryALTinterwordspacing
J.~Chorowski, D.~Bahdanau, D.~Serdyuk, K.~Cho, and Y.~Bengio, ``Attention-based
  models for speech recognition,'' \emph{CoRR}, vol. abs/1506.07503, 2015.
  [Online]. Available: \url{http://arxiv.org/abs/1506.07503}
\BIBentrySTDinterwordspacing

\bibitem{Battenberg}
E.~{Battenberg}, R.~J. {Skerry-Ryan}, S.~{Mariooryad}, D.~{Stanton}, D.~{Kao},
  M.~{Shannon}, and T.~{Bagby}, ``Location-relative attention mechanisms for
  robust long-form speech synthesis,'' in \emph{ICASSP 2020 - 2020 IEEE
  International Conference on Acoustics, Speech and Signal Processing
  (ICASSP)}, 2020, pp. 6194--6198.

\bibitem{guidedloss}
\BIBentryALTinterwordspacing
H.~Tachibana, K.~Uenoyama, and S.~Aihara, ``Efficiently trainable
  text-to-speech system based on deep convolutional networks with guided
  attention,'' \emph{CoRR}, vol. abs/1710.08969, 2017. [Online]. Available:
  \url{http://arxiv.org/abs/1710.08969}
\BIBentrySTDinterwordspacing

\bibitem{LJS}
K.~Ito and L.~Johnson, ``The lj speech dataset,''
  \url{https://keithito.com/LJ-Speech-Dataset/}, 2017.

\bibitem{mcd}
R.~{Kubichek}, ``Mel-cepstral distance measure for objective speech quality
  assessment,'' in \emph{Proceedings of IEEE Pacific Rim Conference on
  Communications Computers and Signal Processing}, vol.~1, 1993, pp. 125--128
  vol.1.

\bibitem{libritts}
\BIBentryALTinterwordspacing
H.~Zen, R.~Clark, R.~J. Weiss, V.~Dang, Y.~Jia, Y.~Wu, Y.~Zhang, and Z.~Chen,
  ``Libritts: A corpus derived from librispeech for text-to-speech,'' in
  \emph{Interspeech}, 2019. [Online]. Available:
  \url{https://arxiv.org/abs/1904.02882}
\BIBentrySTDinterwordspacing

\bibitem{li2019jasper}
J.~Li, V.~Lavrukhin, B.~Ginsburg, R.~Leary, O.~Kuchaiev, J.~M. Cohen,
  H.~Nguyen, and R.~T. Gadde, ``Jasper: An end-to-end convolutional neural
  acoustic model,'' 2019.

\end{thebibliography}
